\documentclass[a4paper,superscriptaddress,twocolumn,amsmath]{revtex4}
\usepackage{amsfonts}
\usepackage{euscript}
\usepackage{graphicx}
\usepackage[latin1]{inputenc}
\usepackage{amsmath}
\usepackage{amssymb}
\usepackage[latin1]{inputenc}
\usepackage{amsmath}
\usepackage{amssymb}
\usepackage{amsfonts}
\usepackage{graphicx}
\usepackage[latin1]{inputenc}
\usepackage{amsmath}
\usepackage{amssymb}

\begin{document}


\title{An angular rainbow of light from curved spacetime.} 
\author{Alexei A. Deriglazov }
\email{alexei.deriglazov@ufjf.edu.br} 
\affiliation{Depto. de Matem\'atica, ICE, Universidade Federal de Juiz de Fora,
MG, Brazil,} \affiliation{Department of Physics, Tomsk State University, Lenin Prospekt 36, 634050, Tomsk, Russia}

\date{\today}

\begin{abstract}
We try to go beyond the geometrical optics approximation, by showing that a massless polarized particle allows a wide class of non minimal interactions with an arbitrary gravitational field. One specific example of a curvature-dependent interaction is presented, that results in a frequency-dependent Faraday effect. Even in a Schwarzschild spacetime, this leads to the angular dispersion of polarization planes for a linearly-polarized beam of waves with different frequencies, propagating along the same ray.
\end{abstract}

\maketitle 


Description of light rays in a gravitational field beyond the geometrical optics approximation has a long history \cite{Skrotskii_1957, Plebanski_1960}, and is widely discussed in the current literature \cite{Frolov_2011, Andrey_Shoom_2020, Frolov_2020, Oancea_2020, Terno_2021, Chakraborty_2021, Hamada_2018, Nucamendi_2020, Yunlong_Liu_2020, Sheoran_2020, Carapulin_2020}.  It is expected, that account of polarization and chirality of the light beam could lead to a number of dynamic effects, similar to those that occur for spinning particles in a gravitational field \cite{Yu_Peng_Zhang_2020_1, Yu_Peng_Zhang_2019_2, Biswas_2020_1, Toshmatov_2020, Azreg-Ainou_2020, Mukherjee_2020, Bin_Chen_2020, Kaye_Jiale_Li_2020, Brax_2, Du_2021, Benavides_2021, Compere_2021, Walberto_2021, Bofeng_Wu_2021, Ke_Yang_2021, Atamurotov_2021, Jiale_Li_2021, Battista_Li_2021}. First, influence of spin-orbit interaction on the trajectory is under discussion \cite{Frolov_2011, Yamamoto_2018,  Dolan_2018, Harte_2019, Armen_2020, Andrey_Shoom_2020, Oancea_2020, Frolov_2020}, and is considered as a gravitational analogue of Magnus or spin Hall effects of light observed in medium \cite{Zeldovich_1992}. Second, dynamics of polarization plane is under the interest \cite{Skrotskii_1957, Plebanski_1960, Ishihara_1988, Fayos_1982, Nouri-Zonoz_1999, Yihan_Chen_2011, Connors_1980, Sereno_2004, Ghosh_2016, Schneiter_2018}. This is considered as a gravitational analogue of the classical Faraday effect. Importantly, concerning the gravitational Faraday rotation discussed in the literature, it does not depend on the wave frequency. Besides, it is due to $g_{0i}$\,-components  of the metric, so the Faraday rotation is not expected in Schwarzschild space.

Various approaches have been used to theoretically describe and analyze the polarization degrees of freedom and related effects \cite{Bialynicki_1996, Kosinski_2016, Mohrbach_2006, Dodin_2015_1, Skrotskii_1957, Plebanski_1960, Frolov_2011}.  In particular, when studying the propagation of light outside the geometrical optics approximation \cite{Frolov_2011, Andrey_Shoom_2020, Frolov_2020,  Dolan_2018, Yamamoto_2018}, a special solution of the Maxwell equations in curved spacetime is sought, resembling, as close as possible, a plane wave in flat space.  With this solution can be associated a kind of a massless particle, endowed with extra degrees of freedom describing the wave polarization or helicity. The form of resulting equations of motion depend on the anzatz, that was chosen for the special solution. In particular, it is possible to obtain the equations with the trajectory depending on a helicity, that is, predicting the gravitational spin-Hall effect. 

In this paper we use a somewhat opposite way: starting from a plane wave in flat space, we can associate with it a massless polarized particle, and then construct the corresponding variational problem. Within the variational problem, we can try to construct an interaction of the particle with gravity. Following this way, we show that the minimally interacting particle turns out to be equivalent to the Maxwell equations in the geometrical optics approximation. Then a nonminimal interaction can be thought as an alternative way to go beyond the geometrical optics approximation.  

The aim of this paper is to develop, in a systematic form, the manifestly covariant Lagrangian description of a massless polarized particle (photon, for short) in curved spacetime.  We propose one version of such a coarse-grained model of the photon, and see to what extent it is able to capture the properties of light propagation in curved spacetime.

{\bf Lagrangian of a massless polarized particle in curved space-time.}\label{lagrangian}\label{ss3}
First we show that basic equations of the geometrical optics can be obtained as the conditions of extreme of the following variational problem:
\begin{eqnarray}\label{lag.7}
S=\int d\tau ~ \frac{1}{2\tilde e_1}\left[g_{\mu\nu}+\tilde e_3\frac{\omega_\mu\omega_\nu}{\omega^2}\right]Z^\mu Z^\nu+\frac{1}{2\tilde e_2}(\nabla\omega)^2, 
\end{eqnarray}
where $Z^\mu=\dot x^\mu+\tilde e_4\nabla\omega^\mu$.
We work with the ray trajectory $x^\mu(\tau)$ in an arbitrary parameterization $\tau$.  The variable $\omega^\mu(\tau)$ is a mechanical
analogue of the vector potential $A^\mu(x^\nu)$ of electromagnetic field, $\nabla\omega^\mu=\dot\omega^\mu +\Gamma^\mu{}_{\rho\sigma}\dot x^\rho\omega^\sigma$ is the covariant derivative for the metric $g_{\mu\nu}(x^\rho)$, and $\tilde e_i$ are the auxiliary variables. In the spinless limit, $\omega^\mu\to 0$, the functional (\ref{lag.7}) reduces to the standard Lagrangian action of a massless point particle, $\int d\tau\frac{1}{2e}\dot x^2$. We note also that 
the action (\ref{lag.7}) does not represent a massless limit of the massive spinning particle \cite{AAD_Rec}.

Replacing $g_{\mu\nu}$ in (\ref{lag.7}) by the Minkowski metric  $\eta_{\alpha\beta}$, we obtain the manifestly Poincare invariant action with two Noether charges. They are the conjugated momentum for $x^\alpha$, and the total angular momentum: $p_\alpha$, $ J_{\alpha\beta}=x_{[\alpha}p_{\beta]}+ \omega_{[\alpha}\pi_{\beta]}$. Here $p_\alpha=\frac{\partial L}{\partial\dot x_\alpha}$ and $\pi_\alpha=\frac{\partial L}{\partial\dot\omega^\alpha}$. Note that $x_{[\alpha}p_{\beta]}$ and $\omega_{[\alpha}\pi_{\beta]}$  are not preserved separately. Note also that $p^\alpha$ and $\pi^\alpha$ are not proportional to $\dot x^\alpha$ and $\dot\omega^\alpha$. 

To study the theory (\ref{lag.7}), we use the Hamiltonian formalism, which is well adapted for the analysis of  a constrained theory \cite{Dirac_1977, Gitman_1990, AAD_Book}. Conjugate momenta for $\omega^\mu$ are denoted as $\pi_\mu=\frac{\partial L}{\partial\dot\omega^\mu}$, and the covariant canonical momenta for $x^\mu$ are ${\cal P}_\mu=\frac{\partial L}{\partial\dot x^\mu}-\pi_\sigma\Gamma^\sigma{}_{\mu\rho}\omega^\rho$. The Hamiltonian form of our variational problem, $S_H=\int d\tau ~ p\dot x+\pi\dot\omega-H$, is much more transparent, showing that the Hamiltonian 
\begin{eqnarray}\label{lag.12}
H=\frac{e_1}{2}{\cal P}^2+\frac{e_2}{2}\pi^2+e_4{\cal P}\pi+\frac{e_3}{2\omega^2}({\cal P}\omega)^2, 
\end{eqnarray}
is a combination of the Dirac constraints 
${\cal P}^2=0$,  ${\cal P}\pi=0$, $\pi^2=0$,  ${\cal P}\omega=0$.   
Using the tetrad formalism \cite{Landau_2}, we conclude that the constraints surface consist of four subsurfaces. Only on the subsurface 
\begin{eqnarray}\label{sub}
\pi_\mu=0, \quad {\cal P}^2=0, \quad {\cal P}\omega=0,
\end{eqnarray}
our equations of motion are self-consistent (in other sectors, they either imply $\dot x^0=0$, or the number of equations is more than the number of independent variables).  
Even in the sector (\ref{sub}), $\omega^\mu$ is unobservable quantity because of its dynamics is not causal.  But it can be used to construct an observable quantity: 
$f_{\mu\nu}={\cal P}_\mu\omega_\nu-{\cal P}_\nu\omega_\mu\equiv{\cal P}_{[\mu}\omega_{\nu]}$.
For the variables $x^\mu$, ${\cal P}^\mu$ and $f_{\mu\nu}$, the variational problem (\ref{lag.12}) in the sector (\ref{sub}) implies the equations
\begin{eqnarray}\label{lag.24}
\dot x^\mu=e_1{\cal P}^\mu, \qquad \nabla{\cal P}^\mu=0, \qquad \nabla f_{\mu\nu}=0, 
\end{eqnarray}
where $\nabla$ is the covariant derivative, $\nabla{\cal P}^\mu=\dot{\cal P}^\mu+\Gamma^\mu_{\rho\sigma}\dot x^\rho {\cal P}^\sigma$. 
The dynamical equations are accompanied by the algebraic relations
\begin{eqnarray}\label{lag.25}
{\cal P}^2=0, \qquad f_{\mu\nu}{\cal P}^\nu=0, \qquad \tilde f_{\mu\nu}{\cal P}^\nu=0,
\end{eqnarray}
implied by the basic constraints ${\cal P}^2={\cal P}\omega=0$.  By $\tilde f$ we denoted the tensor dual to $f$. 

So, we confirmed that physical sector of the Lagrangian theory (\ref{lag.7}) coincides with that of Maxwell equations, taken in the geometrical optics approximation, the equations (\ref{lag.24}) and (\ref{lag.25}). Denote $E_i\equiv f_{i0}$, $B_i\equiv\frac12\epsilon_{ijk}f^{jk}$. In the Minkowski space the constraints (\ref{lag.25}) imply, that ${\bf p}$, ${\bf E}$ and ${\bf B}$ are mutually orthogonal and ${\bf B}=[\hat{\bf p}, {\bf E}]$, so the triad $(\hat{\bf p}\equiv{\bf p}/p^0, {\bf E}, {\bf B})$ is the right-handed, and moves with the speed of light in the direction ${\bf p}$. This, in essence, is our massless polarized particle. By construction, it can be used to study the dynamics of electric and magnetic fields of a plane monochromatic wave  along a chosen ray (null geodesic). 

The remaining ambiguity due to $e_1$, contained in the equations (\ref{lag.24}), is due to their reparametrization invariance. It disappears in the coordinate-time parameterization, $\tau=t$, when we exclude ${\cal P}^0$ with use of the constraint ${\cal P}^2=0$. By doing this, we get the equations 
\begin{eqnarray}\label{lag.30}
\frac{d{\bf x}}{dt}=c\boldsymbol{\hat{\cal P}}, \quad \nabla_t\boldsymbol{\hat{\cal P}}-c\boldsymbol{\hat{\cal P}}\Gamma^0{}_{\mu\nu}\hat{\cal P}^\mu\hat{\cal P}^\nu=0,  \quad \nabla_t\tilde\omega=0,\label{lag.31} \\
\nabla_t f_{\mu\nu}=0, \quad f_{\mu\nu}\hat{\cal P}^\nu=\tilde f_{\mu\nu}\hat{\cal P}^\nu=0. \label{lag.33}
\end{eqnarray}
We denoted $\hat{\cal P}^\mu=(1, \boldsymbol{\hat{\cal P}})$. The variables $\hat{\cal P}^1, \hat{\cal P}^2, \hat{\cal P}^3$ are subject to the constraint  
${\bf g}\boldsymbol{\hat{\cal P}}+\sqrt{\boldsymbol{\hat{\cal P}}\gamma\boldsymbol{\hat {\cal P}}}/\sqrt{-g_{00}}=1$. Here ${\bf g}$ and $\gamma$ are components of $(3+1)$\,-decomposition of the original metric \cite{Landau_2}: $g_i=-\frac{g_{0i}}{g_{00}}$, $\gamma_{ij}=g_{ij}-\frac{g_{0i}g_{0j}}{g_{00}}$. 
In the limit of flat space, the variables $\tilde\omega$ and $\hat{\bf p}$ are related with the original variables as follows: $\hat{\bf p}={\bf p}/|{\bf p}|$, $\tilde\omega=c|{\bf p}|$. So, the unit vector $\hat{\bf p}$ is normal to the wave front, while $\tilde\omega$ can be identified with  the frequency of the wave, that represents our particle. This  fixes the dimension of canonical momenta of a massless particle: $[{\cal P}^\mu]=1/cm$. 

{\bf Frequency-dependent Faraday effect in Schwarzschild spacetime due to a non minimal interaction.}\label{ss5}
The inclusion  of an interaction, preserving the physical sector of a free theory with Dirac constraints, generally represents a non trivial task. In our case, the interaction must be introduced so that the (deformed) constraints still admite the subsurface (\ref{sub}) as one of the solutions. It is remarkable, that this condition is fulfilled when the term $H_{int}=e_1\pi_\mu\Omega^\mu(x, {\cal P}, f)$,  
with any function $\Omega$  orthogonal to ${\cal P}$
\begin{eqnarray}\label{non.0}
{\cal P}_\mu\Omega^\mu=0, 
\end{eqnarray}
is added to the Hamiltonian (\ref{lag.12}). In the sector (\ref{sub}) the theory remains consistent, and equations of motion are (\ref{lag.24}) and (\ref{lag.25}), except the dynamical equation for $f$, that now reads
\begin{eqnarray}\label{non.5}
\nabla f_{\mu\nu}=e_1{\cal P}_{[\mu}\Omega_{\nu]}.
\end{eqnarray}
Hence the parallel transport of $f$ is disturbed by the non minimal interaction.

Let us discuss a number of specific examples of covariant couplings. If we restrict ourselves with the linear on curvature and polarization interactions,  the terms with desired property (\ref{non.0}) are $\Omega^\mu\sim R^\mu{}_{\nu\rho\delta}{\cal P}^\nu \tilde f^{\rho\delta}$, and $\Omega^\mu\sim R^\mu{}_{\nu\rho\delta}{\cal P}^\nu f^{\rho\delta}$. Choosing the term with dual tensor $\tilde f^{\mu\nu}$, the interaction term in the Hamiltonian is
$H_{lin}=\tilde\kappa e_1 \pi_\mu R^\mu{}_{\sigma\rho\delta}{\cal P}^\sigma \tilde f^{\rho\delta}$, 
where $\tilde\kappa$ is a coupling constant. In the coordinate-time parameterization, Eq. (\ref{non.5}) reads 
\begin{eqnarray}\label{non.8}
\nabla_t f_{\mu\nu}=\tilde\kappa\tilde\omega\hat{\cal P}^\varphi g_{\varphi[\mu}R_{\nu]\sigma\rho\delta}\hat{\cal P}^\sigma \tilde f^{\rho\delta},
\end{eqnarray}
where, $\tilde\omega=c{\cal P}^0$ represents the frequency of the photon, and $\hat{\cal P}^\sigma={\cal P}^\sigma/{\cal P}^0$. 
We assumed that ${\cal P}^\mu$ of the  massless particle has the dimension of a wave vector, $[{\cal P}^\mu]=1/cm$. Then the coupling constant has the dimension 
$[\tilde\kappa]=1/cm^2$. Combining the dimensional constants at our disposal, we can write $\tilde\kappa=l_P^2\kappa$, where $\kappa$ is already dimensionless, and 
$l_P=\sqrt{\frac{\hbar G}{c^3}}$ is the Planck length. The linear interaction then will be very small, being suppressed by square of the Planck length $\sim 10^{-66}$ cm. To improve this, we try the non linear interactions. We can try to divide $H_{lin}$ on any one of the following expressions: $\left(R_{\sigma\lambda\mu\nu}R^{\sigma\lambda\mu\nu}\right)^{\frac12}$, 
$\left(\nabla_\sigma R^{\sigma}{}_{\lambda\mu\nu}\nabla_\delta R^{\delta\lambda\mu\nu}\right)^{\frac13}$, 
$\left(\nabla_\delta R_{\sigma\lambda\mu\nu}\nabla^\delta R^{\sigma\lambda\mu\nu}\right)^{\frac13}$, $\ldots$ , all them are of dimension $1/cm^2$. The interaction, constructed with help of $\sqrt{R^2}$, does not vanish in the limit of plane space $M\rightarrow 0$, so we reject it. The second expression  is not appropriate, since it vanishes on-shell due to the Bianchi identity: $\nabla_\sigma R^{\sigma}{}_{\lambda\mu\nu}=\nabla_\mu R_{\lambda\nu}-\nabla_\nu R_{\lambda\mu}=0$. Using the third expression, we have the interaction
\begin{eqnarray}\label{non.8.1}
H_{int}=\kappa e_1 \pi_\mu R^\mu{}_{\nu\rho\delta}{\cal P}^\nu \tilde f^{\rho\delta}\left(\nabla_\delta R_{\sigma\lambda\mu\nu}\nabla^\delta R^{\sigma\lambda\mu\nu}\right)^{-\frac13}, 
\end{eqnarray}
where $\kappa$ is a dimensionless coupling constant. This implies the following equation of motion for $E_i=f_{i0}$
\begin{eqnarray}\label{non.9}
\frac{dE_i}{dt}=c\Gamma^\sigma{}_{\rho [0}\hat{\cal P}^\rho f_{i]\sigma}+
\kappa\tilde\omega\frac{\hat{\cal P}^\varphi g_{\varphi[i}R_{0]\sigma\rho\delta}\hat{\cal P}^\sigma \tilde f^{\rho\delta}}{(\nabla R, \nabla R)^{\frac13}}.
\end{eqnarray}
The right hand side of this equation is a sum of torques due to minimal and non minimal interactions. 
It shows that parallel transport of  polarization vector is disturbed by space-time curvature. Besides, contrary to the case of minimal interaction, the frequency $\tilde\omega$  now entered into the equation for ${\bf E}$. To understand the consequences of this new property, we analyze this equation in the leading order approximation, in Schwarzschild metric taken in the Cartesian isotropic coordinates \cite{Landau_2}
\begin{eqnarray}\label{sc.2}
g_{\mu\nu}dx^\mu dx^\nu=-\frac{(1-\frac{\alpha}{4|{\bf x|}})^2}{(1+\frac{\alpha}{4|{\bf x|}})^2}(dx^0)^2+(1+\frac{\alpha}{4|{\bf x|}})^4d{\bf x}d{\bf x}.
\end{eqnarray}
We denote $\alpha=2MG/c^2$,  $|{\bf x}|=\sqrt{x^i x^i}$ and $\hat x^i=\frac{x^i}{|{\bf x}|}$. Note that $\Gamma$ and $R$ of this metric are of order $1/c^2$.  So, computing the leading contributions into  Eq. (\ref{non.9}), we can use the approximation $g_{\mu\nu}=\eta_{\mu\nu}$ for all other quantities presented in this equation. Then the algebraic relations (\ref{lag.25}) imply, that  the set $(\hat{\boldsymbol{\cal P}}, {\bf E}, {\bf B})$ is a triad of mutually orthogonal vectors. We are interested in the dynamics of  polarization plane, that is we need to compute the component of the vector $d{\bf E}/dt$ in the direction of ${\bf B}$.  The direct calculation gives the following result: 
\begin{eqnarray}\label{sc.9}
\left.\frac{d{\bf E}}{dt}\right|_B=-\kappa\tilde\omega\EuScript{C} ~ \sqrt[3]{\frac{12MG}{5c^2|{\bf x}|}} 
[\hat{\boldsymbol{\cal P}}, {\bf E}]\equiv[{\boldsymbol{\Omega}}, {\bf E}],
\end{eqnarray}
where $\EuScript{C}\equiv 1-(\hat{\bf x},\hat{\boldsymbol{\cal P}})^2-2(\hat{\bf x}, \hat{\bf B})^2$.  
As it should be \cite{Plebanski_1960, Ishihara_1988, Fayos_1982, Nouri-Zonoz_1999, Yihan_Chen_2011},  all $1/c$\,-terms of Eq. (\ref{non.9})), originated from the minimal interaction, do not contribute into Eq. (\ref{sc.9}), and the Faraday rotation in Schwarzschild field is exclusively due to the non minimal interaction (\ref{non.8.1}). 

Eq. (\ref{sc.9}) has clear interpretation for the typical scattering process, when an ingoing at $t=-\infty$ polarized particle is propagated towards the region of Schwarzschild field, and then it is observed in the asymptotically Minkowski region at $t=+\infty$. Eq. (\ref{sc.9}) states that  the axis ${\bf E}$ precess around the vector ${\boldsymbol{\Omega}}\sim\hat{\boldsymbol{\cal P}}$ with the angular velocity 
$|{\boldsymbol{\Omega}}|=\kappa\tilde\omega|\EuScript{C}|\sqrt[3]{12MG/5c^2|{\bf x}|}\sim 1/c^{\frac23}$.  Note that it is more in magnitude as compared with the angular velocity due to $\Gamma$\,-terms in Kerr space \cite{Ishihara_1988, Nouri-Zonoz_1999}. An ingoing linearly polarized at $t=-\infty$ wave, entering into the region with non vanishing curvature, will experience the Faraday rotation, and appeared at $t=+\infty$ with the polarization plane rotated with respect to that of at $t=-\infty$. 

Consider the case of ingoing linearly polarized wave of given frequency, and  with the vector ${\bf E}$ on the plane of motion. Then $(\hat{\bf x}, \hat{\bf B})\approx 0$ during all scattering process, so the angular function $\EuScript{C}$ grows starting from $0$ up to $1$ when ${\bf x}\rightarrow {\bf x}_p$, and then decreases up to $0$ when ${\bf x}\rightarrow \infty$. Here ${\bf x}_p$ is the point of perihelion, that is the point of closest approach to the Schwarzschild center. This means that precession accumulates during the evolution. In particular, this might be an important effect in the study of photon spheres near the horizon of black holes \cite{Toshmatov_2020, Nucamendi_2020, Yunlong_Liu_2020, Sheoran_2020}. 

If the vector ${\bf E}$ of ingoing photon is orthogonal to the plane of motion, we have $-2(\hat{\bf x}, \hat{\bf B})=-2$ at the point of perihelion. Then the angular function has an opposite sign: $\EuScript{C}({\bf x}_p)=-1$, and the vector ${\bf E}$ rotates in the opposite direction, as compared with the previous case. 

According to Eq. (\ref{sc.9}), the rotation angle linearly depends on the wave frequency $\tilde\omega$. 
So the Schwarzschild spacetime play a role of dispersive media for the polarization axes of the waves with different frequency. Consider the linearly polarized light beam composed of waves with different frequencies, but with the same polarization axis at $t=-\infty$. They will follow the same trajectory,  but leave the region with a different orientation of the polarization axes for waves with different frequency, forming the angularly distributed rainbow, produced by the non minimal polarization-curvature interaction (\ref{non.8.1}). 

In conclusion, two comments are in order. \par 
\noindent 
1. According to the modifications beyond the geometrical optics approximation \cite{Oancea_2020, Frolov_2020}, curvature-dependent interactions could be responsible for the gravitational spin-Hall effect.  Concerning the gravitational Faraday effect, it requires one to take into account the next-to-leading order approximation \cite{Fayos_1982, Ishihara_1988, Nouri-Zonoz_1999}, and hence involve derivatives of the connection.  If these contributions can be written in a covariant form, we expect that they also should depend on the curvature. So, our simple model with non minimal curvature-dependent interactions suggests an alternative way for analysis of these effects beyond the leading order of geometrical optics. \par 
\noindent   

2. By construction, our model does not take into account the helicity of a photon, so we can not expect that it could capture the spin-Hall effect. However, it is interesting that the Hall-type term appears in the equation for ${\cal P}$: $\nabla{\cal P}_\mu=R_{\mu\nu\rho\delta}\dot x^\nu\pi^\rho\omega^\delta$, as an inevitable consequence of the covariantization: $\dot \omega^\mu\rightarrow\nabla\omega^\mu$. According to the recent works \cite{Frolov_2011, Frolov_2020, Oancea_2020}, namely such a term could be responsible for the gravitational spin-Hall effect of light. But in our model it vanishes in the region (\ref{sub}) of the constraints surface, where the theory is a self-consistent.
One possibility to improve this point is to relax the constraints, assuming a non standard dispersion relation instead of ${\cal P}^2=0$, as it was suggested in \cite{Frolov_2011, Oancea_2020, Frolov_2020}. However, it is not clear, how to do this in a way, consistent with the requirement of coordinate independence of the speed of light in general relativity.

{\bf Acknowledgments.} 
The work has been supported by the Brazilian foundation CNPq,  and by Tomsk State University Competitiveness Improvement
Program.

\end{document}